\documentclass[12pt,preprint]{aastex}
\def\kms{{\rm km\,s^{-1}}}

\def\masyr{{\rm mas}\,{\rm yr}^{-1}}

\def\lim{{\rm lim}}

\def\nom{{\rm nom}}
\def\true{{\rm true}}

\begin{document}

\title{An Upper Limit on the Granularity of the Local Stellar Halo}

\author{Andrew Gould}
\affil{Department of Astronomy, The Ohio State University,
140 W.\ 18th Ave., Columbus, OH 43210}
\and
\affil{Institute for Advanced Study, Princeton, NJ, 08540}
\authoremail
{gould@astronomy.ohio-state.edu}

\singlespace

\begin{abstract}

I use the statistical properties of 4588 nearby halo 
stars to show that if the local stellar halo is composed of cold streams 
from disrupted dwarf satellites as is predicted by popular cosmologies, 
then at 95\% confidence there are at least 1350 such streams.  Moreover,
no stream can contain more than 2.3\% of the local stars, while
average streams must contain less than 0.08\%.

\end{abstract}
 
\section{Introduction
\label{sec:intro}}

In the conflict between the two competing models for the formation of
the Milky Way, the \citet{sz} scenario of cannibalistic assimilation
of pre-existing protogalaxies has generally been gaining ground
on the earlier monolithic collapse model of \citet{els}.  In part this
is because the former fits naturally into the picture of hierarchical 
structure formation predicted by cold dark matter models, which 
have so many successes on larger scales.  Even more compelling is
the direct observational evidence.  The best example is the
Sgr dwarf galaxy \citep{igi}, whose disruption is taking
place right before our eyes.  However, while it is especially 
striking (\citealt{maj} and references therein), 
the Sgr dwarf is by no means unique.  Palomar 5 \citep{oden,rockosi} and
possibly the Ursa Minor dwarf spheroidal \citep{palma}
show extended tidal tails as evidence of ongoing disruption.

However, these disintegrating satellites are all in the outer
halo.   While individual moving groups have been reported in the
local stellar halo for several decades \citep{eggen}, 
it is not clear whether these
associations are an exotic froth on an underlying well-mixed homogeneous
stellar population, or whether the local stellar halo
is basically granular, being composed of a number of streams, each
tightly clumped in velocity space but so extended in physical
space that they cannot be distinguished photometrically from the
myriad of other streams.  

Numerical simulations demonstrate that kinematically cold, initially 
spherical systems will form cold streams when tidally disrupted
\citep{mcglynn}, and such disruption has been studied in considerable
detail \citep{johnston}.  \citet{bkw} simulated the formation of a
network of streams within the framework of 
hierarchical structure formation and showed that halo tracers such as
RR Lyrae stars would exhibit dramatic variations in surface
density projected on the sky.  While these variations are most pronounced
at large distances, they are also clearly discernible in the simulated
solar neighborhood.  Simulations of satellite disruption by \citet{hw}
also support the granular picture.  They specifically predict a local halo 
composed of 300--500 streams, each with a velocity dispersion of order 
$5\,\kms$.  

Here I show that if the local stellar halo is composed of such 
streams, they are extremely numerous, with typical streams 
each contributing less than 0.08\% to the local stellar halo.

\section{Limits on Granularity
\label{sec:limits}}

In \citet{halo}, I analyzed a sample of 4588 halo stars selected
from the revised NLTT catalog (rNLTT) \citep{faint,bright} by means
of an optical-infrared reduced proper motion (RPM) diagram \citep{rpm}.
In particular, I found that five of the nine parameters describing
the local halo velocity ellipsoid were consistent with zero.   These
were the mean radial and vertical motions of the halo relative to
the local standard of rest (LSR),
$V_1 = 1.4\pm 2.2\,\kms$ and $V_3=1.8 \pm 2.4\,\kms$,
and the normalized off-diagonal components
components of the velocity dispersion tensor,
$\tilde c_{12} = 0.024\pm 0.014$, 
$\tilde c_{13} = 0.005\pm 0.023$, and
$\tilde c_{23} = -0.004\pm 0.026$.  (Of the remaining four velocity-ellipsoid 
parameters, the fit values of the three velocity dispersions
were found to be compatible with previous determinations and the
motion in the tangential direction relative to the Sun was fixed 
as part of the analysis at the previously determined value of
$U_2=-216.6\,\kms$.)

There are many possible effects that could have driven these five numbers
away from zero, which is the value predicted by the simplest axisymmetric
Galactic model.  For example, if the LSR were moving inward or outward
relative to the Galactic potential, the reflex motion of the halo
would have driven $V_1$ away from zero.  Or if the halo had been
rotating around the 2-axis, this would have shown up in $V_3$.
Or if the dark halo were significantly triaxial with principle
axes misaligned relative to standard Galactic coordinates, this would
have driven $\tilde c_{12}$ to significant non-zero values.
However, and this is the key point, there are no systematic
effects that could have driven these parameters systematically toward
zero if that were not their actual value.  In particular, the likelihood
routine actually fit for motions relative to the Sun, which were found
to differ from zero by 5.2 and 2.3 sigma.  It was only after the
known motion of the LSR was removed that they became consistent with
zero.  Since these five parameters agree with the predictions of the simplest
model, their statistical fluctuations, as codified by $\chi^2$, can be used
to test the statistical properties of the underlying sample.
Note that $\chi^2_\nom=3.97$ for 5 degrees of freedom (dof).

Let us now suppose that the local halo were composed of 459 streams, each
with 10 stars and each with an internal velocity dispersion well below
the $100\,\kms$ to $170\,\kms$ values measured for the three components 
of the halo.  The 10 stars would then
have had essentially identical kinematics, and I would therefore have
overcounted the number of independent measurements by a factor of 10.
The best fit parameters for the halo that I derived would have been
correct, but the errors would have been underestimated by a factor of
$10^{1/2}$.  Hence, $\chi^2$ would have been 
$\chi^2_\true = \chi^2_\nom/10 = 0.397$.  This scenario can be
immediately ruled out because the probability of finding such a low
$\chi^2$ for 5 dof is less than 0.5\%.

More generally, one can imagine that the 4588 stars in the sample
are divided into $N$ streams, with the $i$th stream having $m_i$
members, $m_i\geq 1$.  Then $\chi^2$ would have been overestimated
by $\sum_i m_i^2/4588$. Hence,
\begin{equation}
\chi^2_\true = {4588\over \sum_i m_i^2} \chi^2_\nom.
\label{eqn:chi2true}
\end{equation}
I now demand that $\chi^2_\true>1.145$, so that it does not fall within
the 5\% lowest part of the $\chi^2$ distribution for 5 dof. I
thereby obtain a limit,
\begin{equation}
\sum_{i=1}^N m_i^2 =  {\chi^2_\nom\over\chi^2_\true}\sum_{i=1}^N m_i < 15908,
\qquad \sum_{i=1}^N m_i = 4588.
\label{eqn:sumlimit}
\end{equation}
If the $m_i$ were real variables, the minimum value of $N$ that
would saturate this constraint would be $N=4588^2/15908=1323$,
with $m_i=15908/4588=3.47$.  Since the $m_i$ must be integers,
the minimum is $N=1350$, in which case there would be 816
streams with 3 stars and 535 with 4 stars.  That is, each stream
would have about 0.08\% of the total local halo population.
The maximum number of stars that can be in any stream is
106.  In this case, all the other 4483 ``streams'' would have only
one star each.

\section{Discussion
\label{sec:discuss}}

If the statistical errors in \citet{halo} had been overestimated, this
would have artificially decreased $\chi^2$ and so made the limits
derived in the present {\it Letter} appear stronger than they actually are.  
Is such an error overestimation a possibility?  No.  The \citet{halo} errors
are strongly dominated by Poisson statistics, not measurement errors,
so that even if the measurement errors were set to zero in the analysis, 
the errors on the derived parameters would not be reduced by more than 
a few percent \citep{pg}.  

Suppose that the \citet{halo} halo sample were contaminated at a fractional
level $f$ by another population (e.g.\ thick disk stars) whose
five velocity-ellipsoid parameters were also zero.  Suppose that the
halo sample were clumpy but the other population was not.  The lower
limit derived in \S~\ref{sec:limits} should then be decreased by
a fraction $f(r-1)/(r-f)\sim 0.7f$, where 
$r\equiv \chi^2_\nom/\chi^2(5\%)=3.47.$  From figure 3 of \citet{faint},
I estimate $f\sim 2\%$, so this effect is negligibly small.

Can the technique presented here be extended to probe more effectively
for granularity in the stellar halo?
The rNLTT covers only 44\% of the sky and moreover is derived from
a proper-motion selected catalog $\mu>180\,\masyr$.  If the catalog
were extended to most of the rest of the sky (see \citealt{faint}),
the number of stars could be doubled.  By comparing first epoch
Palomar Observatory Sky Survey (POSS I) plates with Early Release
Data from the Sloan Digital Sky Survey (SDSS), \citet{digby} were able
to identify halo stars from a RPM diagram for proper motion stars down to 
$40\,\masyr$.  This extends the
distance out to which they can be found by a factor 4.5 relative to 
rNLTT and therefore increases the volume by a factor 90.  Because
halo stars are distributed roughly uniformly in space, the sample size
will increase in proportion to the volume.
Since SDSS will eventually
cover 1/4 of the sky, the prospects for increasing the sample
by a factor $\sim 50$ are good.  Future surveys may do even better.
If the five velocity ellipsoid parameters continue to be consistent
with zero, it will be possible to push down the limits on halo
granularity by roughly the same factor.

But what if some of these parameters begin to deviate significantly
from zero?  How can one tell, within the framework of this approach,
whether these formally significant detections reflect the real properties
of the halo velocity ellipsoid or an underestimation of the velocity-ellipsoid
errors due to clumpiness of the stellar halo?

Two of the five parameters, $V_1$ and $\tilde c_{12}$ are much more
likely to really be non-zero than the other three.  If the halo
is triaxial and one of its axes is not perfectly aligned with the
Sun, then this should induce both inward (or outward) net motion
of the LSR ($V_1$) and a vertex deviation in the halo ($\tilde c_{12}$).  
If one or both
of these parameters begins to differ significantly from zero, but
the remaining three remain consistent with zero, then the same
test developed here can be used to constrain granularity, except
that it would have to be restricted to the three remaining parameters,
$V_3$, $\tilde c_{13}$, and $\tilde c_{23}$.  
On the other hand, if halo granularity is causing the errors to be 
underestimated, it should affect both groups of parameters equally.

While the statistical power of a $\chi^2$ test for 3 dof is somewhat
weaker than for 5 dof, this would be more than compensated by the
increase in star counts.  Thus, there are reasonable prospects for
simultaneously carrying out high-precision tests of Galactic structure
using two of the five parameters while tightly constraining or precisely
measuring the degree of halo granularity using the other three.

The limits derived here are in strong formal conflict with the
predictions of \citet{hw}.  No formal comparison can be made
with the predictions of \citet{bkw} because they do not report on
the quantity measured here, the total number of cold streams in
the local halo.  It is nevertheless possible that a prediction for
this quantity could be extracted from their simulation.

However, this formal conflict does not necessarily contradict the
picture of hierarchical structure formation on galaxy scales.
The simulations conducted to date have considered one or a few
realizations without an attempt to tune parameters to fit any
quantitative measurement of streams.  With the advent of the limits
presented here, as well as the prospect of much better limits and/or
measurements in the reasonably near future, it should be possible 
to quantitatively constrain such scenarios or, perhaps, to rule
them out.

\acknowledgments 
I thank Scott Tremaine and Scott Gaudi for stimulating discussions.
This work was supported by grant AST 02-01266 from the NSF and by
JPL contract 1226901.


\clearpage


\end{document}